\documentclass{article}

\usepackage{arxiv}
\usepackage{array}
\usepackage[utf8]{inputenc} % allow utf-8 input
\usepackage[T1]{fontenc}    % use 8-bit T1 fonts
\usepackage{hyperref}       % hyperlinks
\usepackage{url}            % simple URL typesetting
\usepackage{booktabs}       % professional-quality tables
\usepackage{amsfonts}       % blackboard math symbols
\usepackage{nicefrac}       % compact symbols for 1/2, etc.
\usepackage{microtype}      % microtypography
\usepackage{lipsum}
\usepackage{graphicx}
\graphicspath{ {./images/} }

\title{ProblemChild: Discovering Anomalous Patterns based on Parent-Child Process Relationships}

\author{
 Bobby Filar \\
  Elastic \\
  \texttt{filar@elastic.co} \\
   \And
 David French \\
  Elastic \\
  \texttt{david.french@elastic.co} \\
}

\begin{document}
\maketitle
\begin{abstract}
It is becoming more common that adversary attacks consist of more than a standalone executable or script. Often, evidence of an attack includes conspicuous process heritage that may be ignored by traditional static machine learning models. Advanced attacker techniques, like “living off the land"  that appear normal in isolation become more suspicious when observed in a parent-child context. The context derived from parent-child process chains can help identify and group malware families, as well as discover novel attacker techniques. Adversaries chain these techniques to achieve persistence, bypass defenses, and execute actions. Traditional heuristic-based detections often generate noise or disparate events that belong to what constitutes a single attack.  ProblemChild is a graph-based framework designed to address these issues. ProblemChild applies a supervised learning classifier to derive a weighted graph used to identify communities of seemingly disparate events into larger attack sequences. ProblemChild applies conditional probability to automatically rank anomalous communities as well as suppress commonly occurring parent-child chains. In combination, this framework can be used by analysts to aid in the crafting or tuning of detectors and reduce false-positives over time. We evaluate ProblemChild against the 2018 MITRE ATT\&CK™ emulation of APT3 attack to demonstrate its promise in identifying anomalous parent-child process chains.

\end{abstract}

\section{Introduction}
To meet the evolving threat of adversary techniques, Traditional Anti-Virus (AV) leveraged signatures to capture known malicious binaries. Over time this created an arms race between attacker and defender that saw most AVs augment signatures with machine learning in hopes of generalizing previously observed malware against the unknown and Next-Generation Anti-Virus (NGAV) was born. The use of ML to detect and prevent malicious binaries has seen a meteoric rise within the security industry and has become table stakes for endpoint protection platforms. The relative success of this approach at stopping novel attacks with “zero-day” malware infections becoming less commonplace has caused sophisticated adversaries to shift their tactics and techniques.
 
In June 2017 a new ransomware variant called Petya\cite{symantec} was first observed in the wild as it hit a variety of organizations worldwide (primarily focused on Ukrainian companies). Petya leverages an SMB vulnerability made popular by the ransomware family WannaCry. Unlike WannaCry, Petya leveraged a series of system-level commands to gain a foothold and establish persistence. These commands provide a “benign” capability to perform credential dumping, execution of itself, setting a scheduled task, and wiping logs. Petya performed all of these tasks using the tools provided by the operating system like Windows Management Instrumentation (WMI) command-line tool \textit{wmic.exe} and task scheduler tool \textit{schtasks.exe}. Tactics like these are called “Living off the land”\cite{istr2017} and represent the next great challenge in securing the endpoint as attackers rely heavily on dual-use software to execute their attacks and “hide in plain sight” on the operating system.

We present ProblemChild a graph-based framework designed to address the concerns listed above. ProblemChild ingests process Event Tracing for Windows data (ETW) and exposes an intuitive interface to address technical challenges in the detection of living off the land techniques, specifically anomalous or rare parent-child relationships. Our four main contributions are as follows:
\begin{itemize}
\item Training a machine learning model on a large set of malicious and benign event data targeting features from process creation events.
\item Providing an anomaly score that accounts for the prevalence of a parent-child process pair within the local environment against the output of the ML model. This score becomes the edge weight between a parent and child node in a directed graph.
\item We apply community detection to segment the weighted graph into smaller “chains” of processes.
\item Finally, we apply a threshold against each community to determine if it's overall score is anomalous and return a ranked list of potentially malicious communities.
\end{itemize}

We demonstrate that the proposed approach can reduce a large set of process-related events to a manageable list of rank-ordered rare parent-child process chains, suppressing commonly occurring, but previously unobserved activity. Ranking or prioritizing of events has been shown to help reduce noise in an environment\cite{milajerdi2018holmes}.

\section{Background}
\subsection{Living off the Land Binaries}
Living off the Land Binaries \cite{lolbas} are Microsoft-signed binaries that come pre-installed on the operating system. These pieces of software have alternative or unexpected features outside of their core functionality. For example, the binary \texttt{schtasks.exe} allows an administrator to create, delete, query, change, run, and end scheduled tasks on a local or remote computer \cite{stevewhims}. However, this binary may also be leveraged by an attacker to bypass User Account Control (UAC) and escalate privileges (e.g., as in Fig. \ref{fig1}). The use of these binaries reduces the number of new files dropped to disk during an attack and lessens the chance of being detected by security software.

\begin{figure*}

\begin{verbatim}
    reg add HKCU\Environment /v windir /d  \ 
        "cmd /K reg delete hkcu\Environment /v windir /f && REM "
    
    schtasks /Run /TN \Microsoft\Windows\DiskCleanup\ SilentCleanup /I
\end{verbatim}
\caption{Using \texttt{schtasks.exe} for disk cleanup UAC bypass is an example of a living-off-the-land technique.}\label{fig1}
\end{figure*}

Attackers may also chain these binaries together to perform more sophisticated actions that mimic traditional adversary attack sequences \cite{istr2017}.
\begin{itemize}
    \item \textbf{Incursion}: Initial access vector either exploiting a Remote Code Execution (RCE) vulnerability or targeted individuals using a spear-phishing attack.
    \item \textbf{Persistence}: Post-compromise actions to allow the attacker to maintain a presence on the system.
    \item \textbf{Payload}: Employs dual-use tool (e.g. psexec or mimikatz) or fileless capability to execute rest of attack
\end{itemize}

The use of these binaries makes the discovery of an attack much more difficult as adversary behavior is mixed in with traditional benign operating system activity. Historically, APT groups and other threat actors were identifiable based on the custom binaries they left behind. By using benign OS binaries attribution becomes that much more difficult. After all, these binaries will not be detected by AV engines and there is no exploit taking place. For security vendors, this means exploring alternative data streams like ETW and attempting to derive anomalous or suspicious event sequences based on file, process, DNS, or security-related events.

\begin{figure}
    \centering
        \begin{tabular}{ | m{3.4cm} | m{3.4cm} | m{3.4cm} | } 
        \hline
        Initial Access & Execution & Persistence \\ 
        \hline
        Privilege Escalation & Defensive Evasion & Credential Access \\ 
        \hline
        Discovery & Lateral Movement & Collection \\ 
        \hline
        Command \& Control & Exfiltration & Impact \\ 
        \hline
        \end{tabular}    
    \caption{MITRE ATT\&CK Enterprise Tactics}
    \label{fig:fig2}
\end{figure}
\subsection{MITRE ATT\&CK™ Framework}
The industry has responded to the living off the land threat by rallying around frameworks like MITRE ATT\&CK. The MITRE ATT\&CK Framework is a knowledge base that seeks to describe adversary behavior by providing a standardized taxonomy of attack and defense security techniques (e.g. See Fig. \ref{fig:fig2}). At the core of the ATT\&CK framework is matrix/ontology that provides a structured visualization of the how and the why of an attack:
\begin{itemize}
    \item  \textbf{Tactic}: presentation of an adversary’s objective or reason for taking an action.
    \item  \textbf{Technique}: tactical objective of taking an action
    \item  \textbf{Group}: Known adversaries that have carried out a specified technique
    \item  \textbf{Software}: The binary used to execute a specified technique
\end{itemize}

Mass adoption of the ATT\&CK framework is in large part due to the attention given to parent-child relationships as a detection methodology for living off the land techniques. This methodology is further enhanced by providing conditionals based on metadata associated with the parent-child process chain (e.g. observing a MS Office process spawn powershell.exe with base64 encoded arguments).

\subsection{Detection Engineering}
The current solution to combat these techniques revolves around developing a rule or heuristic to act as a detector for a given technique. Threat researchers have moved to simple, schema independent query languages to craft real-time detectors against streaming data.\cite{endgameinc_2019} Language like EQL allows researchers to craft expressive rule-based logic for detecting a specific ATT\&CK technique.  Fig. \ref{fig:fig3} shows an example rule to detect scriptable child processes of Office products and a Spearphishing Attachment (T1193) technique\cite{spearphishing_mitre}.

\begin{figure*}
    
    \begin{verbatim}
    process where
        parent_process_name in ("winword.exe","excel.exe","powerpnt.exe")
        and process_name in ("powershell.exe","cscript.exe", "wscript.exe","cmd.exe")
    \end{verbatim}
    \caption{Example EQL rule for (T1193)}
    \label{fig:fig3}
\end{figure*}

This approach is adequate in practice, but does have several limiting factors: Rules can be prone to false positives because they are based on inherently benign software, 2) That noise can contribute to alert deluge common in security software, and 3) The generation of rules is a manual process that requires domain expertise and the ability to learn from or across customer environments to tune detectors over time.

\section{Related Work}
The use of system-level entities, like parent-child process chains, have historically been used to by security tools to model APT kill chains \cite{milajerdi2018holmes}, establish provenance graphs \cite{gehani2012spade}\cite{ma2016protracer}, or construct dependency graphs for root cause detection \cite{milajerdi2018propatrol}.
 
Additionally, there has been research on correlating alert data to reduce alert noise, such as alert management \cite{milajerdi2018holmes} and improving the quality and efficacy of alerts \cite{hassan2019nodoze}. However, applying these approaches in a real-world operation setting can be difficult. Vendors must account for industry, customer, administrator, and user behavior when rolling out a solution. Without careful consideration of environmental factors (e.g. \textit{what is normal in organization X?})  these detectors may become overly sensitive to a specific attack signature and thus be prone to false-positives.
 
When applying detectors based on human-derived logic, customization, in any form, requires a monitoring period and an ability to tune a detector. If these detectors are too specific, some attacks will be captured, but they will fail to generalize to previously unseen or novel variants. An iterative tuning approach suggested by [12] has proven to be a valuable method for developing powerful detectors.
 
However, the approach above is only made possible by having detection content to write rules against. Frameworks like Atomic Red Team\cite{redcanaryco_2019}, RTA \cite{endgameinc_2018}, and Calder \cite{mitre_2019} provide a library of scripts that replicate living off the land techniques described in the MITRE ATT\&CK matrix. These frameworks can be executed by technique ID (e.g. T1088 - Bypass User Account Control) or, in the case of Atomic Red Team, chained together to replicate real-world adversaries. Additionally, there has been a fair amount of work by these researchers to provide a blue team (\textit{defensive}) counterpart to their red team ATT\&CK frameworks. AtomicBlue, MITRE CAR, Windows Defender ATP Queries have all released repositories of detectors or heuristics that map to ATT\&CK techniques in hopes of fostering a community of sharing. While these frameworks are invaluable to threat researchers writing detectors they may also be leveraged as a labeled dataset for machine learning applications.

\section{Approach}
ProblemChild relies on system-level data, primarily process events, extracted from Microsoft ETW data to identify anomalous parent-child process chains and hunt for living off the land attacks. In this study, we restrict our attention to Windows events, but the approach may similarly be applied to Linux (using \textit{auditd}). We import the data in bulk using Sysmon\cite{russinovich2009sysinternals}, an open-source collection tool. 

\subsection{Data Ingest and Normalization}
Data transformation into a numeric representation serves two functions: 1) it helps reduce resource overhead and complexity in storing supplemental data leaving primary data (e.g. \textit{process name}) intact for graphing operations and 2) allows the ProblemChild analytic engine to learn broader details of parent-child relationships, in the scope of an attack, which avoids just learning signatures.
 
Since ProblemChild is a graph-based analytic framework extracted metadata from each process creation event is stored in the following format and stored in graph format:
\begin{itemize}
    \item \textbf{Node} - object or entity being modeled (e.g. \textit{process name})
    \item \textbf{Edge} - action taken by an object (e.g. \textit{process create, fork, terminate})
    \item \textbf{Metadata} - properties or artifacts that describe a node or edge (e.g. \textit{process ID, command line arguments, timestamps})
\end{itemize}
The graph is a directed acyclic graph as employed in other works \cite{mitre_2019,pei2016hercule,shu2015unearthing}.

\subsection{Detecting Anomalous Events}
Having formed a graph, a series of statistical methods are applied to draw out anomalous parent-child chains and attempt to connect disparate detections based on community detection. Community detection seeks to segment a graph structure based on the relative edge weights between nodes. Often this is represented as the number of “links” between two nodes, but for our use-case, we choose a technique similar to \cite{pei2016hercule} to provide weight assignments for each edge via supervised learning. Specifically, we use \emph{XGBoost}\cite{chen2016xgboost}, an implementation of a gradient boosted trees model. Unlike previous research, ProblemChild focuses on learning from past malicious and benign process chains to determine a weight correlated with the maliciousness of a given parent-child pair. This becomes a supervised learning problem, that targets the following features to generate a weight for a given edge between \textit{(u, v)}:
\begin{itemize}
    \item $\Delta t$ between process creation and termination
    \item one-hot encoding of u.child, v.parent\\ (\textasciitilde100 windows processes + 1 \emph{UNK} slot)
    \item Process signature info
    \item Process elevation info
    \item Process integrity info
    \item Is process running as system
    \item Parent-child User mismatch
    \item Entropy of process name
    \item Entropy of command line
    \item TF-IDF (n-grams) of the command line arguments
\end{itemize}

These data points provide a simple feature vector that can be passed to a \emph{XGBoost} model that has been trained on labeled malicious and benign edges derived from event data. Instead of predicting a binary label, we choose to instead predict the class probability between 0.0-1.0 for a malicious label and use that number as the weight for a given edge.

We then run Louvain community detection on the weighted graph \cite{blondel2008fast} to segment the graph. Community detection should aid in the identification of rare process chains (e.g. \textit{attack sequences}) and provide a structure for "grouped" attack techniques. The resulting dictionary consists of node assignments (\textit{e.g. unique processes}) to a given community.

\subsection{Prevalence Engine}
When performing a post-mortem on event data, or in hunting within an environment, it may be useful to understand how prevalent a process, parent-child chain, or a process-command line is within a local or target environment. ProblemChild constructs a service at run-time to apply a series of probability calculations to yield a "weight" to multiply against the output of the  \emph{XGBoost} model. 

First, ProblemChild provides a query interface to get the prevalence of a process using the following method:
\begin{center}Pr(process\_count < PERCENTILE)\end{center}
The score will range between 0 and 99 which represents the percentile of the number of times the process has been seen in the wild. The process prevalence score allows ProblemChild to state \textit{"I've seen this process more than x\% of other processes"}. Likewise, detecting the prevalence of a Parent-Child Process chain requires a similar calculation, except we perform conditional probability first, instead of raw counts of a parent-child relationship. So, we let:
\begin{center}P(child | parent) = P(child, parent) / P(parent)\end{center}
Since the output of this equation can range widely, we then apply the same percentile technique as above to ensure 0-99 score. Thus, the parent process prevalence score becomes:
\begin{center}Pr(P(child | parent) < PERCENTILE)\end{center}
The analyst can then determine \textit{"From this parent, I've seen this child more than X\%of other child processes".}  This interface to the parent-child prevalence score is accessible via a dictionary lookup similar to prevalence[child][parent]. 

\begin{figure*}
\begin{verbatim}
def find_bad_communities(G, threshold):
  bad_communities = [ ]
  communities  = community_detection(G, weight=weight)

  for community in communities:
    for node in community:
        prevalence_score = prevalence[node.process_name][node.pprocess_name]
        node['anomalous_score'] = node['weight'] * (1- prevalence_score)
	
    if max([node['anomalous_score'] for node in community]) >= threshold:
        bad_communities.append(community)
  
  # return most malicious communities first
  return sorted(bad_communities, reverse=True)
\end{verbatim}
    \caption{Python code for ranking communities by maliciousness}
    \label{fig:my_label}
\end{figure*}

\subsection{Finding Bad Communities}
We sought to maximize the malicious\_score (\textit{global ML view}) of a given parent-child chain by combining it with its prevalence\_score (\textit{local statistical view}) to generate an anomalous\_score. We performed this calculation for every node in the graph by iterating through each community (see code block). We then perform a max() across all the anomaly scores for nodes in a given community, if the returned value is >= the specified threshold the community is deemed malicious and set aside for user review. Upon completing this action for each community in the graph the list of malicious communities is rank-ordered to return the highest scored communities first and suppressing commonly occurring process chains.  We hope this will drastically reduce the amount of data and threat researchers have to work with and limit the creation of noisy detectors.

\section{Experiments}
Prior to the primary experiment, ProblemChild was trained on multiple datasets (both malicious and benign) to establish a normal, global baseline. As mentioned in the previous section, the classifier was used to generate edge weights. The prevalence service was comprised of data from the experiment dataset, representing a local view of rare/common processes. 

\subsection{Datasets}
The ProblemChild model was trained on a combination of real-world and simulated benign and malicious data. Benign data consisted of 3 days of internal Endgame ETW process data. The sources of this data were a mix of user workstations and servers to replicate a small organization. Then we generated malicious data by detonating all the ATT\&CK techniques available via the Endgame RTA framework, as well as launching macro and binary-based malware from advanced adversaries like FIN7 and Emotet. We used this combined dataset to train the XGBoost model mentioned in the Approach Section.

Similarly, our evaluation data was comprised of 2 additional days of internal Endgame ETW process data (benign) and Atomic Red Team attack data, plus APT3 data that simulates the 2018 MITRE ATT\&CK evaluation (malicious). All data was collected using Windows SysInternals Sysmon\cite{russinovich2009sysinternals}. 

\begin{figure}
    \centering
    \includegraphics[width=3.7in]{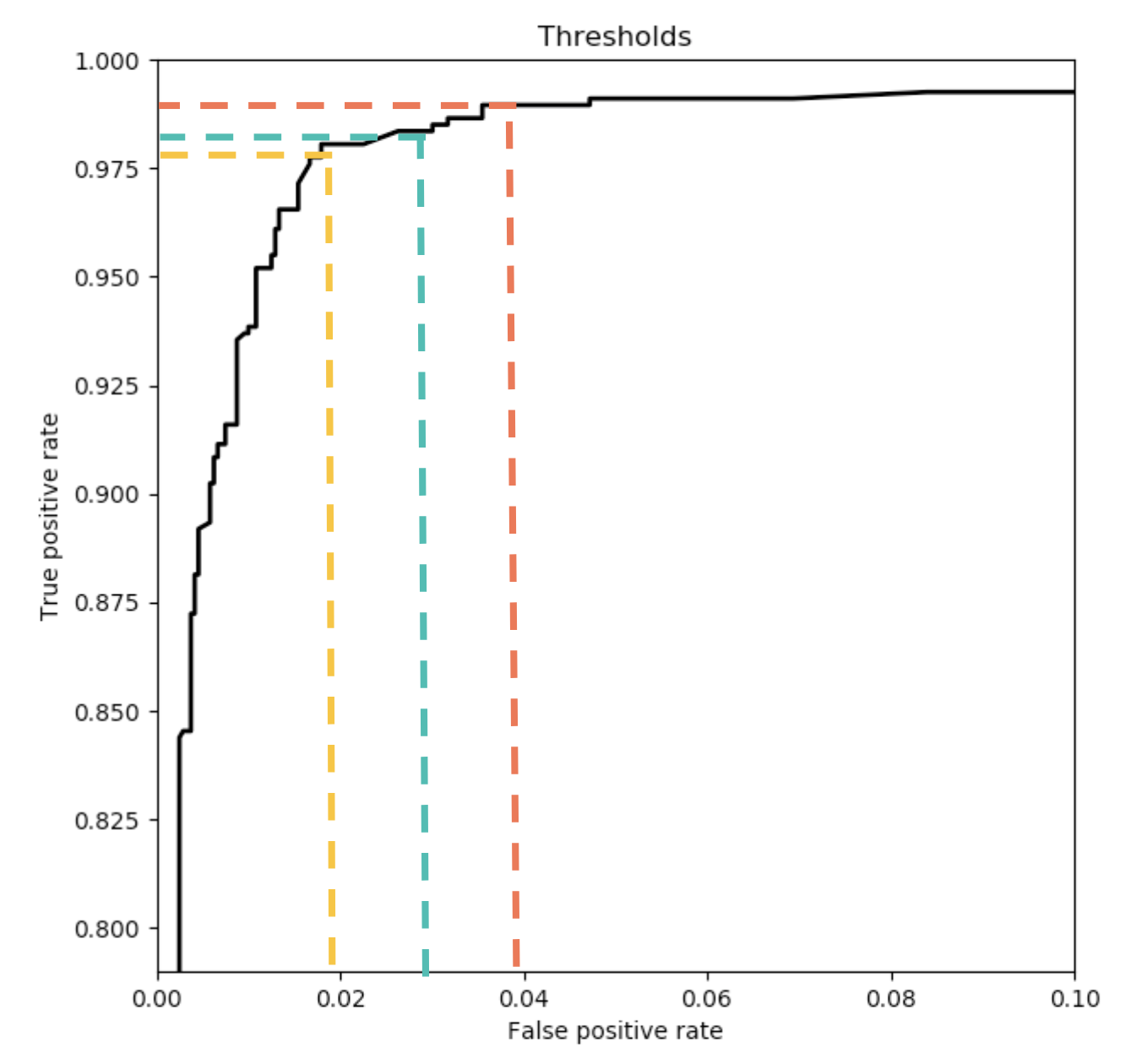}
    \caption{ROC curve w/ target TP/FP Rates}
    \label{fig:rocf}
\end{figure}

\subsection{Determining a Threshold via Leave-one-out (LOO) Retraining}
We performed LOO retraining as a method to determine an ideal threshold of malicious communities. The process consisted of retraining a classifier on all but one subset of the training data and then evaluating the holdout to produce a false-positive rate (FPR) and false-negative rate (FNR) for a given model. For example, we trained the ProblemChild classifier on all but one day of benign data from machine A. We then gathered performance metrics by evaluating the model against the held-out data and moved on to repeat the process against data from machine B. Upon completion of this process we determined the ideal threshold to maintain an FPR < 3\% was 0.38 (Fig. \ref{fig:rocf}). 

The primary cause of false-positives was due to an admin machine that leveraged PowerShell scripting for various tasks (e.g. pushing out updates, automation). Likewise, false-negatives were largely attributed to attack chains that had a single parent-child process event followed by file, registry, and network events.

\subsection{Setup}
The experiment was presented as a blue team post-mortem exercise. In this case, the red team executed a series of sophisticated attacks that melded together multiple techniques from the ATT\&CK matrix. We decided to use event data from the 2018 MITRE ATT\&CK Evaluation provided by Cyb3rWard0g’s Mordor project \cite{hunters-forge_2019}. The ATT\&CK Evaluation sought to emulate APT3 activity using FOSS/COTS tools like PSEmpire and CobaltStrike. These tools allow living off the land techniques to be chained to perform Execution, Persistence, or Defense Evasion tasks.
\subsection{Results – 2018 MITRE ATT\&CK Evaluation (APT3)}
APT3 group represents a Chinese-based threat actor associated with China’s Ministry of State Security\cite{recorded_future_2017}. The primary target(s) of APT3 campaign had been the US, but in mid-2015 the group shifted focus to political organizations in Hong Kong. APT3 was emulated for the 2018 ATT\&CK Evaluation because of its robust, and diverse, post-exploitation tradecraft which relies heavily on living off the land techniques. While this experiment was an ambitious test of ProblemChild’s capabilities there are inherent shortcomings with the setup. First, there is limited user noise in the environment. Second, since the machines were not actively used during the exercise, the adversary activity is unrealistically loud. We still believe it is a valid test of our framework due to the sophistication and diversity of techniques employed.
 
Overall, ProblemChild was able to recognize groups of ATT\&CK techniques executed by the (PowerShell) Empire Project \cite{psempire_2019}. An example can be seen in below:
\begin{verbatim}
node8: {'score': 0.7542
  chain: [
    ('cmd.exe','C:\\Windows\\system32\\cmd.exe /C ipconfig /all & arp -a & echo  
    USERDOMAIN%\\%USERNAME% & tasklist /v & sc query & net start & systeminfo & 
    net config workstation'),           (1)
    ('ipconfig.exe', 'ipconfig  /all'), (2)
    ('arp.exe', 'arp  -a '),            (2)
    ('tasklist.exe', 'tasklist  /v'),   (3)
    ('sc.exe', 'sc  query '),           (4) 
    ('systeminfo.exe', 'systeminfo')    (5)
    ]}

\end{verbatim}

This process chain above contains the following techniques:
\begin{enumerate}
\item System Owner/User Discovery (\textit{T1033})
\item System Network Configuration Discovery (\textit{T1016})
\item Process Discovery (\textit{T1057})
\item System Service Discovery (\textit{T1007}) 
\item System Information Discovery (\textit{T1082})
\end{enumerate}

Likewise, ProblemChild was able to identify sequences of Multi-step Persistence, Execution, Exfiltration attempts and lateral movement. However, ProblemChild struggled against sequences where process events were not the primary trigger. ATT\&CK techniques like Multi-band Communication (\textit{T1026}), Remote Desktop Protocol (\textit{T1076}), and Create Account (\textit{T1136}) were completely missed by the platform, highlighting the importance of incorporating additional events types in future research.

Finally, we were able to demonstrate the value of the Prevalence Engine in suppressing FPs that arose during this scenario and reducing the overall number of process creation events from 50K down to 40 "malicious" communities consisting of \textasciitilde4-6 events in each.

\section{Future Work}
While the overall performance of ProblemChild demonstrated value in the reduction of process events by identifying anomalous parent-child process chains, a known short-coming is the lack of diverse, realistic data. More work is required to execute, capture, and label malicious binaries. Additionally, we must significantly increase the amount of real-world benign data. Data from admins, developers, and normal enterprise users. That data coupled with the inclusion of other event types (e.g. DNS, registry) should greatly improve our efficacy in delineating good/bad process behavior. Feature extractors for each new event type will require input from threat researchers to determine valuable metadata to target for featurization and inclusion into the model. 

Finally, we must continue to better understand the overall performance of ProblemChild in scenarios similar to the APT3 evaluation. We have limited data on the classification performance in this experiment due to the lack of a complete labeled dataset.

\section{Conclusion}
We have presented ProblemChild, a graph-based analysis framework for automatically uncovering anomalous process-level events. The application of local prevalence increases the ability of ProblemChild to hone in on rare parent-child chains and connect them to larger attack patterns.  While detectors based on heuristics perform well in identifying a singular living off the land technique, we believe there is an opportunity for machine learning to help reduce and rank the event data threat researchers based these detectors on, limiting a deluge of FPs.

\bibliographystyle{unsrt}  


\end{document}